\documentclass[aps, pra, a4paper, notitlepage, reprint, amsfonts, amsmath, amssymb,floatfix]{revtex4-1}

\usepackage{amsmath}
\usepackage[utf8]{inputenc}
\usepackage{graphicx}				
\usepackage{epstopdf}
\usepackage{xcolor}
\usepackage{physics}
\usepackage{hyperref}

\begin{document}

\title{Spin-1/2 XXZ Heisenberg chain in a longitudinal magnetic field}
\author{Mykhailo V. Rakov$^{\dag}$}

\author{Michael Weyrauch}
\affiliation{Physikalisch-Technische Bundesanstalt, Bundesallee 100, D-38116 Braunschweig, Germany}
\affiliation{$^{\dag}$Technische Universit\"at Braunschweig, Mendelssohnstra\ss e 3, D-38106 Braunschweig, Germany}
\date{\today}

\begin{abstract}
We study the XXZ Heisenberg model in a longitudinal magnetic field using a tensor renormalization
method. Built into the tensor representation of the XXZ model is the U(1) symmetry,
which is systematically maintained at each renormalization step. This enables rather
large tensor representations. We extract ground state properties as well as the low lying spectrum from the fixed point tensors.
With rather moderate numerical effort
we achieve a very good accuracy as demonstrated by comparison with Bethe Ansatz calculations.
The phase structure of the model can be accurately reproduced just from the largest fixed point tensor elements.
\end{abstract}
\maketitle

\section{Introduction}

Spin models are of interest for two reasons: firstly, they model in a simple
way the magnetic properties of various  crystals and therefore provide
a suitable basis for physical understanding. Secondly,
they are often amenable to rigorous mathematical analysis. A well-known example is the spin-1/2 XXZ Heisenberg chain.
The model is integrable and therefore most of its properties may be obtained exactly.
An elementary survey of the physics of this model may be found in Ref.~\cite{MikKol}.

The mathematical tools required for the analytic solution of a spin model are
often rather advanced and specific for a particular model. However,
not all spin models are integrable, and  numerical tools are required
to gain quantitative insight into the physics of such models.
A famous example for a widely applicable numerical method for the study of spin models is the density matrix renormalization
group (DMRG) which was first applied  to the spin-1 Heisenberg model with
enormous success~\cite{WHI93a, WHI93b}.

Recently, tensor network methods emerged as promising numerical tools to describe
classical and quantum many-body systems. This is based on the fact that
the partition function of a classical statistical system or
the path integral representation of a quantum system may be approximated
by the tensor trace of a tensor network~\cite{LEV2007, GU2009}.
In principle, numerical evaluation of physical quantities just requires the
evaluation of a tensor trace. Obviously, such a
calculation is exponentially hard in two and more dimensions, and it is precisely this issue which is addressed by
tensor renormalization methods, which convert this exponentially
hard problem into a problem which can be solved in polynomial time with
polynomial memory resources. By now there are many different methods and algorithms
which essentially implement the classical
real-space coarse-graining idea of Kadanoff~\cite{Kadanoff_1966}, i.e. a coarse graining iteration until a fixed point tensor
is found. One determines the physical properties of the particular model from the fixed-point tensor, which is (approximately)
invariant under coarse graining.

A first practical implementation was described
by Levin and Nave~\cite{LEV2007} and named `Tensor
Renormalization Group' (TRG). A deeper understanding of the method and its application were
established in Ref.~\cite{GU2009}. A more efficient renormalization method based on the higher order SVD was proposed in
Ref.~\cite{XIE2012} and was named HOTRG. Later
Evenbly and Vidal~\cite{PhysRevLett.115.180405} introduced a  coarse graining method (TNR)
which proved to be closely related to the MERA (multi-scale entanglement renormalization
ansatz)~\cite{PhysRevLett.115.200401} and, therefore, enables particularly precise calculations close to critical points.
TNR shares this advantage with the `loop' algorithm introduced in Ref.~\cite{PhysRevLett.118.110504}.

In the present study we apply the HOTRG method  proposed in Ref.~\cite{XIE2012} to the spin-1/2 XXZ model.
This model shows a rather rich phase structure, and is therefore ideally suited for a test of tensor renormalization
methods. We not only study the ground state properties but also the low-lying spectrum as obtained from the
fixed-point tensors. This enables significantly more detailed investigations and tests than those previously available. Most studies up to now
concentrated either on the classical or quantum Ising model~\cite{LEV2007,XIE2012,PhysRevB.89.075116,PhysRevLett.118.110504} or the phase boundaries only~\cite{GU2009}.
The low-lying spectrum is not usually considered, since precision calculations require large tensor sizes.
In order to handle large tensor sizes
we introduce U(1) symmetric tensors.
The U(1) tensors introduced here differ from previous studies~\cite{SIN11, PhysRevB.93.054417} and cannot be represented as
arrays as briefly explained in the Appendix. This requires a specific implementation of their algebra.

In section~\ref{sec-ten} we briefly review and explain the HOTRG tensor renormalization method
as far as necessary for our application. Technical details of our implementation are given in Appendix~\ref{U1details}.
A discussion of the results we obtain for the XXZ
Heisenberg model is presented in section~\ref{sec-XXZ}.  A brief description of the U1-Tensors
we implement is given in Appendix~\ref{U1Tensors}.

\section{Tensor network renormalization}\label{sec-ten}

\begin{figure*}
\unitlength1cm
\begin{picture}(18,5.)(0,0)
 \put(1,0)  {\includegraphics[width=16cm]{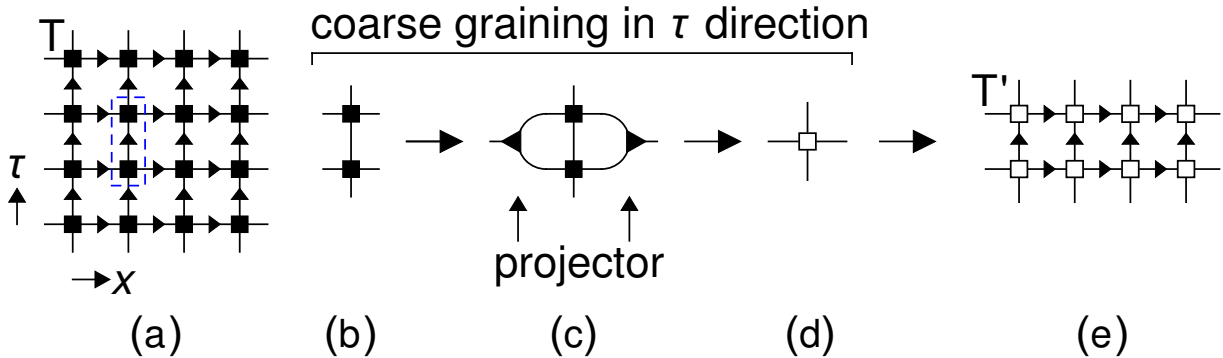}}
\end{picture}
\caption{\footnotesize   The essentials of the HOTRG coarse graining method: (a) a symmetric tensor network as a directed graph. (b) two tensors $T$ symbolized by the small black squares
are contracted into one. (c) unitary projection and approximation using HOTRG.
(d) the renormalized tensor. (e) the coarse grained tensor network.  (Arrows are omitted in (b), (c), (d) for simplicity.)
\label{fig:HOTRG}}
\end{figure*}

We start by expressing the partition function of a 1D quantum system
as a tensor trace
\begin{equation}\label{part func}
Z= {\rm Tr~} e^ {-\beta H}={\rm tTr~}  T^{\otimes K}
\end{equation}
following  Levin and Nave~\cite{LEV2007} and
Gu and Wen~\cite{GU2009}.
Here, $H$ is the Hamiltonian of the many-body system under
consideration and $\beta$ the inverse temperature. The four-index tensors $T_{ijkl}$ are layed out
on a two-dimensional rectangular grid with one time and one space dimension, and the tensor trace includes summation
over all connected indices of the $N$  tensors $T^{\otimes K}$ as illustrated by Fig.~\ref{fig:HOTRG}(a). The (imaginary) time dimension
is discretized into $\tau=\beta/M$ time intervals, and the system size in space direction is $N$ spins, such that $K= N M$ corresponds to the
total number of tensors. We try to achieve as large a grid as possible in order to approximate a system in the thermodynamic limit.

In principle,
there are many different ways to express the partition
sum for a given Hamiltonian as a tensor network. A simple method for Hamiltonians with only
nearest neighbor interactions is described in Refs.~\cite{GU2009, PhysRevB.95.045117}.
We emphasize that the tensors $T$
all have identical structure at each space-time point, i.e. we are dealing with a homogeneous tensor network.
As a consequence, only at most two of them must be stored in computer memory.

Explicit contraction (summation) of the tensor network~(\ref{part func}) in order to calculate the partition sum
is exponentially hard. The network can only be contracted using suitable approximation schemes, often termed
tensor network renormalization (TNR).
Here, we will use an iterative coarse graining  scheme, which
reduces the size of the tensor network by a factor of two at each iteration step. Such methods are related to Kadanoff's block spin idea~\cite{Kadanoff_1966},
and can be implemented rather efficiently.
There are various ways to implement such a scheme, and here we use the higher order SVD (HOSVD)  introduced in
Ref.~\cite{XIE2012}, which provides a very simple way to coarse grain a
tensor network.
The essentials of this method are summarized
in Fig.~\ref{fig:HOTRG}. More details of our implementation are given in Appendix~\ref{U1details}.

Two tensors $T$ are contracted as indicated in Fig.~\ref{fig:HOTRG}(b), then the two left and
the two right indices are contracted with unitary three-leg projectors determined such that the size of the
resulting tensor $T^\prime$ (Fig.~\ref{fig:HOTRG}d) does not increase with respect to the original tensor $T$.
Technically speaking, one approximates the tensor $T$ by the lower-rank tensor $T^\prime$.  The coarse-grained
homogeneous tensor network shown in Fig.~\ref{fig:HOTRG}(e) is made up of these renormalized tensors $T^\prime$.
The higher order (tensor) SVD~\cite{Lat_2000} used for this purpose is a generalization of the standard matrix SVD, which is a well-known tool
in order to approximate a matrix by a lower rank matrix by eliminating small singular values.

After several coarse graining steps, alternating between the space and time directions,
the tensors do change only very little from step to step and we find an approximate fixed point tensor.
From this fixed point tensor we obtain the ground state energy as well as the spectrum rather precisely.
Other properties of the system, e.g. the magnetization, can be calculated as well.
Of course, the actual precision we may achieve in our calculations depends on the size of the dimensions of the tensors we
can handle numerically. This depends on the amount of computer memory available and on the internal structure of the tensors.

Unfortunately, the fixed point tensors we determine are not perfectly stable. If our procedure reaches an approximate fixed point after about 20
iterations alternating between space and time direction, the fixed point destabilizes and the calculation departs from
the physical fixed point for numerical reasons. This corresponds to system sizes of about $30.000\tau$ and 1000 spins at most, however,
such sizes approximate the thermodynamic limit sufficiently for our purposes.
Gu and Wen~\cite{GU2009} pointed out, that the fixed point tensors are contaminated by residual short range entanglement, which they
tried to remove by various entanglement filtering procedures. Here we do not filter out short range entanglement, nevertheless
we find characteristically different fixed point tensors for each quantum phase.

In our particular implementation we take advantage of the fact that all tensors are U(1) symmetric due to the fact that the
XXZ model is U(1) symmetric. The structure of symmetric tensors was elucidated by Singh and Vidal~\cite{PhysRevA.82.050301, SIN11},
who showed that the tensors decompose into a structural and a degeneracy part on the basis of a generalized Wigner-Eckart theorem.
The structural part is given by the symmetry and only the degeneracy part contains the parameters of the tensors. In fact,
only the degeneracy part of symmetric tensors must be stored in memory. In this way we can handle tensor sizes of 130-140 in
each dimension, about a factor 4 larger than in previous calculations~\cite{GU2009}. Details of  U(1) symmetric tensors
are reviewed in Appendix~\ref{U1Tensors}.

\section{XXZ model: ground state energy, gaps and tensor properties}\label{sec-XXZ}

The anisotropic spin-1/2 XXZ model is given by the Hamiltonian
\begin{equation}\label{XXZham}
H=\sum_i  S_x^i  S_x^{i+1}+ S_y^i  S_y^{i+1}+\Delta S_z^i  S_z^{i+1}-\ h S_z^i,
\end{equation}
Here, the $S^i_\lambda$ are spin-1/2 matrix representations of SU(2).
The model depends on two parameters: $\Delta$ (anisotropy) and $h$ (magnetic field);
we will investigate its properties  as a function of these two parameters. Well known special
cases are the Heisenberg model ($\Delta=1, h=0$), and the XX model ($\Delta=0, h=0$).

The XXZ model is U(1) symmetric and its states may be labelled by U(1) quantum numbers $S_z$. Furthermore,
at $h=0$ the model is $Z_2$ spin reflection symmetric, and at the Heisenberg point $(\Delta,h)=(1,0)$ there is an SU(2) symmetry.
These symmetries are reflected in the spectra as well in the phase structure one obtains. However, it is only the
U(1) symmetry which we build explicitly into our numerical algorithm.

The spin-1/2 XXZ Hamiltonian may be diagonalized using the Bethe Ansatz.
For infinite systems, this leads to a Fredholm-type integral equation~\cite{PhysRev.147.303, PhysRev.150.327}, which is easily solved numerically.
For specific parameters (e.g. at the Heisenberg point) the integral equation can be solved analytically.
Results of such calculations are considered to be numerically exact and
serve as convenient benchmarks for our numerical tensor-network investigations. For reference we collect a number of Bethe Ansatz results in Appendix~\ref{appendix-Bethe}.

\begin{figure}
\unitlength1cm
\begin{picture}(18,5)(0,0)
 \put(1,0)  {\includegraphics[width=6.5cm]{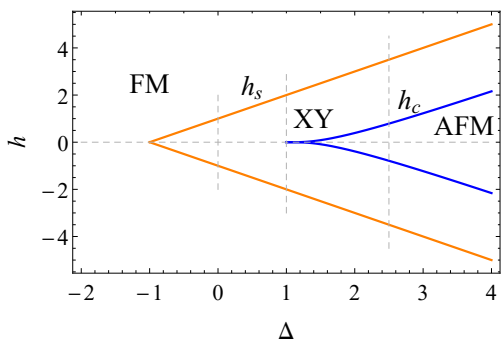}}
\end{picture}
\caption{\footnotesize Phase diagram of the XXZ model in the $(\Delta, h)$-plane. The orange lines separate
the ferromagnetic phase (FM) from the XY phase, and the blue lines separate the XY phase from the
anti-ferromagnetic phase (AFM). The dashed grey lines indicate cross sections in the parameter plane, for which
numerical results will be presented.
\label{fig:phaseDiagram}}
\end{figure}

The phase diagram~\cite{PhysRev.151.258} of the model as obtained using the Bethe Absatz is shown in Fig.~\ref{fig:phaseDiagram}.
We distinguish a ferromagnetic phase (FM), a critically disordered spin liquid phase (XY), and an antiferromagnetic  phase (AFM). These phases are separated by the critical
lines $h_s$ and $h_c$ which start at the critical points $\Delta=\pm 1$.

We now present numerical results obtained with our U(1) symmetric HOTRG code and compare
to corresponding Bethe ansatz results. In particular, we present results for the spectrum and magnetization along the light-grey dashed lines depicted
in the phase diagram~(Fig.~\ref{fig:phaseDiagram}). Strictly speaking, the numerical results correspond to finite size systems of about 500 spins, which are compared to
Bethe ansatz results in the thermodynamic limit. We do not present finite size scaling extrapolations of our numerical results, since such extrapolations are
within the resolution of the presented figures. Typically, the calculations are made with nominal tensor sizes of $m=130$ in each tensor dimension, not counting savings due to U(1) symmetry,
and about 30.000 imaginary time steps with step size $\tau=0.002$. There are more imaginary time steps than spins, because we initially perform 6 HOTRG renormalizations in the imaginary time direction only.
\begin{figure*}
\unitlength1cm
\begin{picture}(18,9)(0,0)
 \put(1,4.5)  {\includegraphics[width=6.5cm]{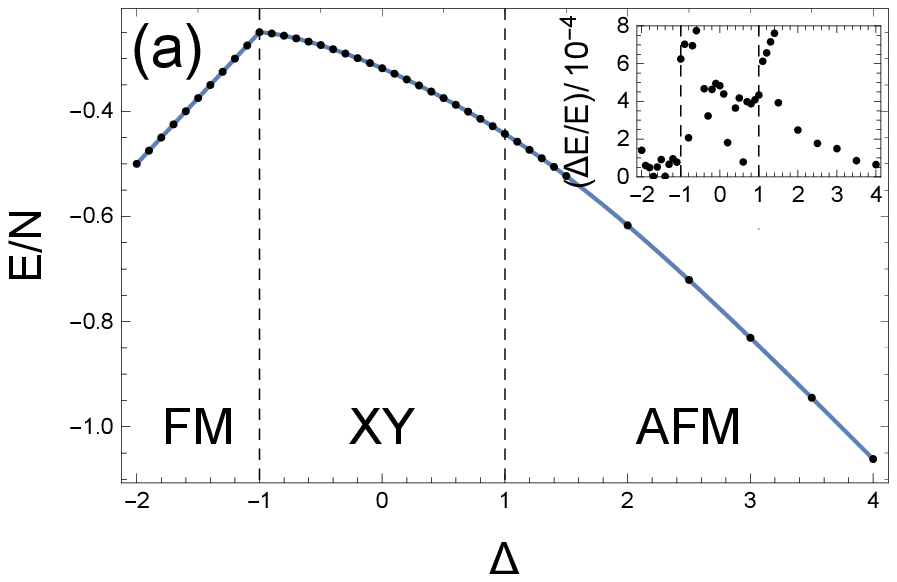}}
 \put(1,0)    {\includegraphics[width=6.5cm]{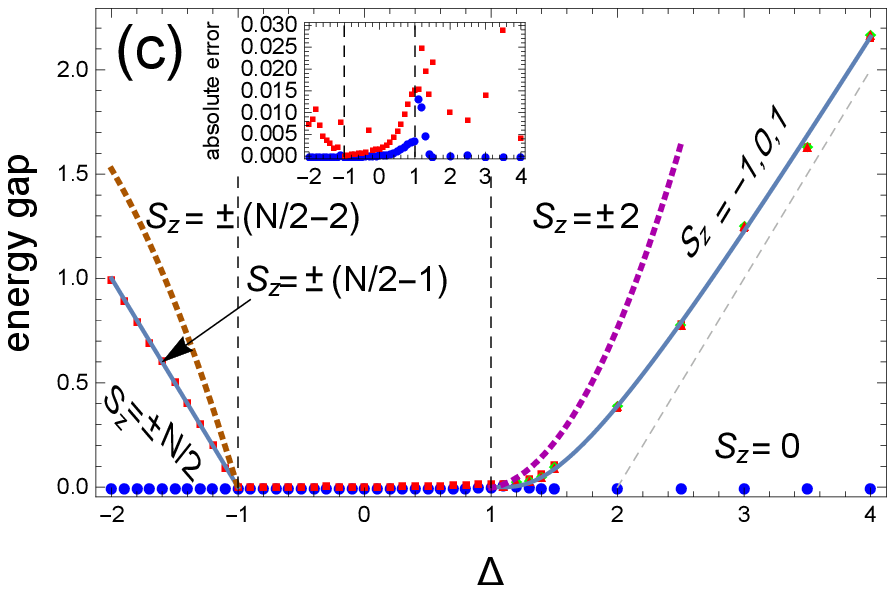}}
 \put(9,0)    {\includegraphics[width=6.5cm]{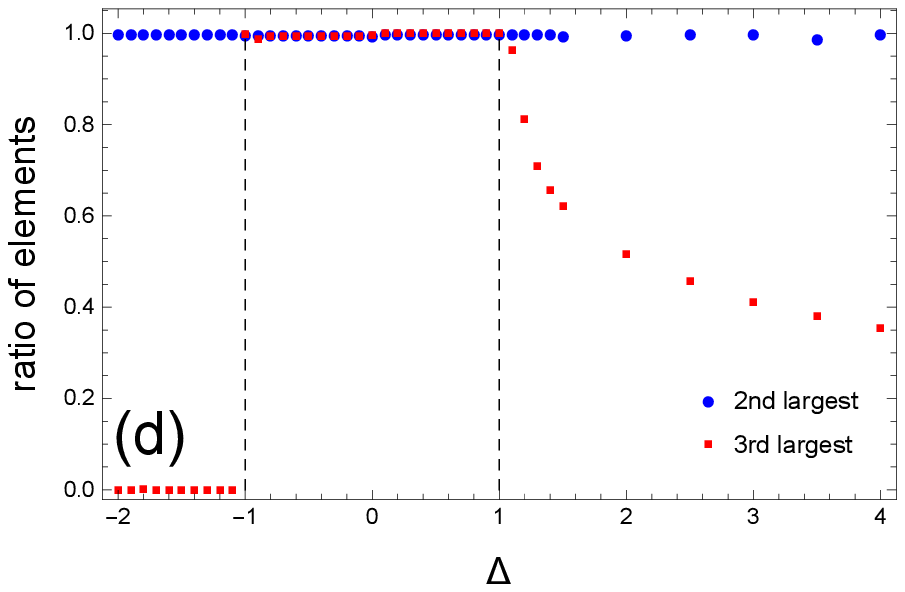}}
 \put(9,4.5)  {\includegraphics[width=6.5cm]{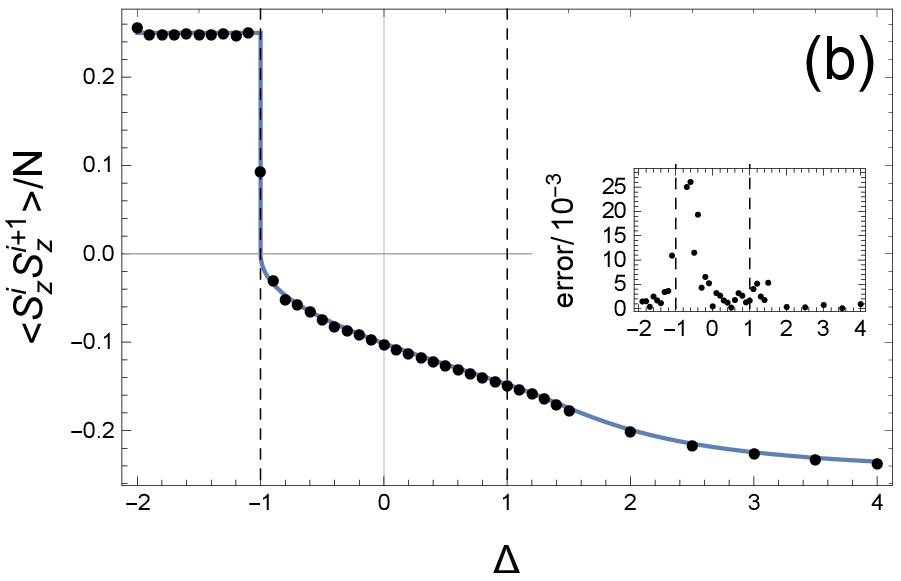}}
\end{picture}
\caption{\footnotesize Properties of the spin-1/2 XXZ model as  functions of the anisotropy $\Delta$ at $h=0$.
(a) ground state energy per site. Full line: analytical result, dots: HOTRG simulations.
Inset: Comparison of the exact ground state energy with HOTRG results.
(b) the spin correlator $\langle S_z^i S_z^{i+1}/N \rangle$. Full line: analytical result, dots: HOTRG simulations. Inset: comparison of exact values with HOTRG results.
(c) spectrum with respect to the ground state energy. Full line: Bethe ansatz, dots: HOTRG simulations, short dashed: asymptotic $\Delta-2$ for the triplet excitations. Inset: absolute errors of the spectral gaps.
(d) absolute values of largest elements of the fixed-point tensor $T$. The largest element is always normalized to 1 and is not shown. The phase boundaries are obtained with high precision from this data.
\label{fig:gseXXZ}}
\end{figure*}

The ground state energy at $h=0$ is shown in Fig.~\ref{fig:gseXXZ}(a) as a function of the anisotropy $\Delta$.
The difference of the numerical results to Bethe ansatz calculations is illustrated in the inset. This difference includes finite size, finite (imaginary) time as well as truncation contributions.
Such effects often contribute with opposite sign and (partly) cancel each other. A detailed analysis of such effects is beyond the present paper.
Not surprisingly, our results indicate that numerical calculations in the critical XY phase are the most difficult. But still, the relative difference $\Delta E/E$ is of the order of $10^{-4}$ over the whole parameter range.
\begin{figure*}[t]
\unitlength1cm
\begin{picture}(18,9)(0,0)
 \put(1,4.5)  {\includegraphics[width=6.5cm]{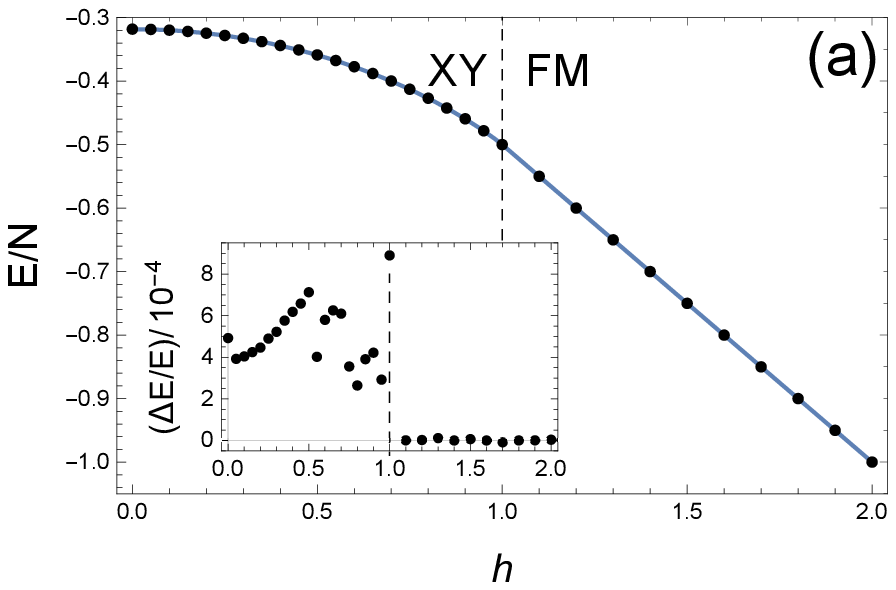}}
 \put(1,0)    {\includegraphics[width=6.5cm]{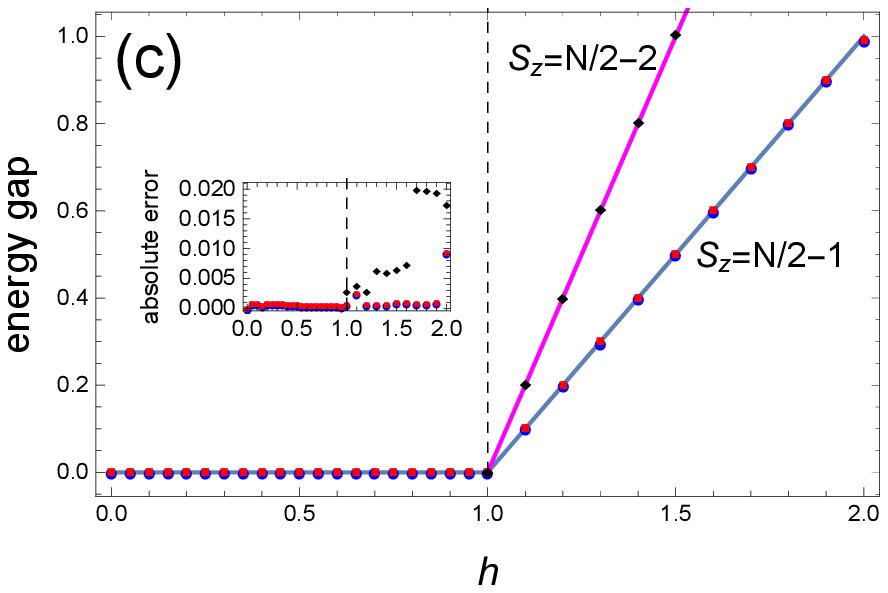}}
 \put(9,4.5)  {\includegraphics[width=6.5cm]{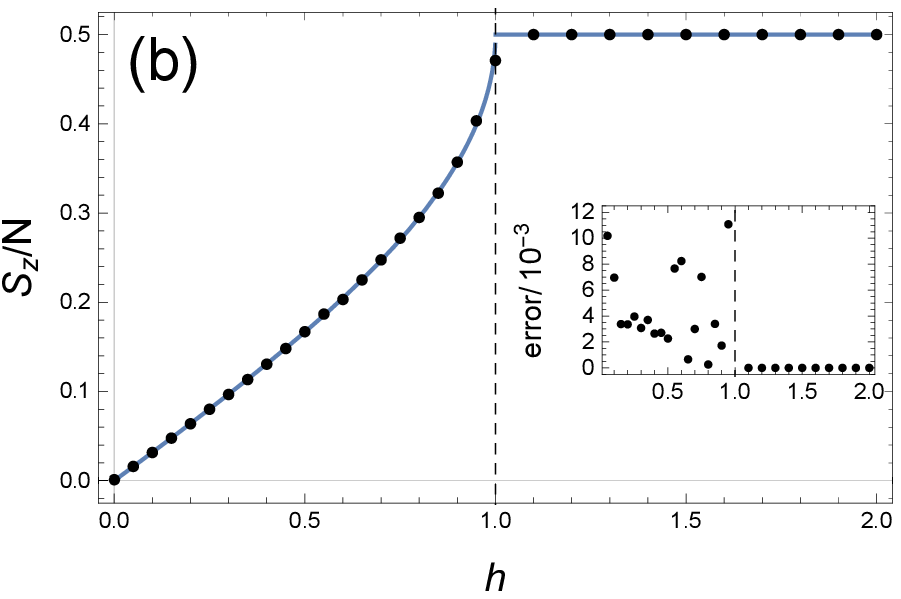}}
 \put(9,0)    {\includegraphics[width=6.5cm]{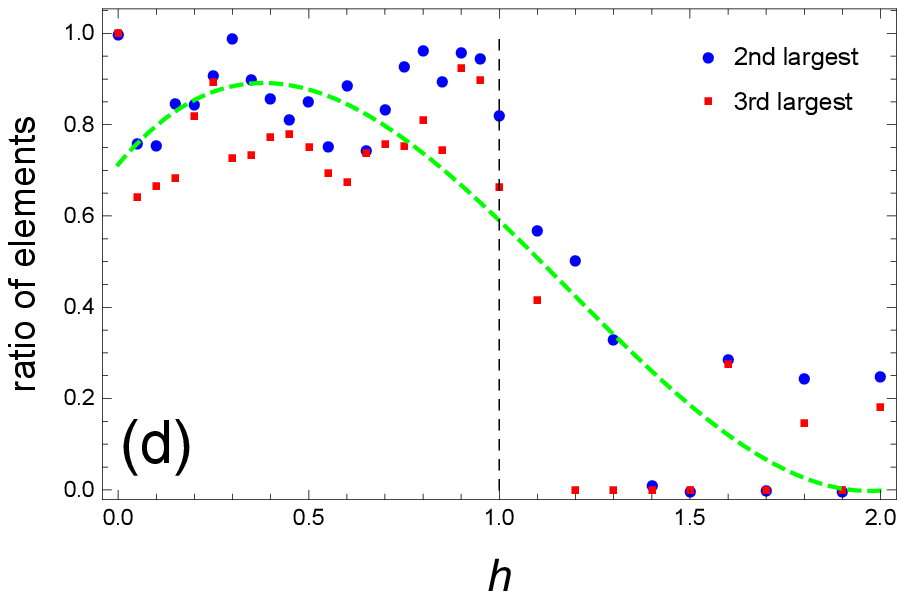}}
\end{picture}
\caption{\footnotesize  Properties of the spin-1/2 XXZ model at $\Delta=0$ and $h\geq 0$: Bethe ansatz (full lines) and HOTRG simulations (dots).
One clearly identifies two phases (XY and FM)  with a second order phase
transition at  $h_{s}=1$.
   (a) ground state energy per site.
              Inset: Comparison of the exact ground state energy with HOTRG results.
   (b) magnetization parallel to the magnetic field.
   Inset: comparison of exact magnetization with HOTRG results.
   (c) gaps to the lowest lying excited states.
   Inset: absolute errors of the spectral gaps.
   (d) Absolute values of the largest elements of the fixed point tensor. The largest element is normalized to 1 and is not shown. The phase boundary can be obtained as inflection point of a polynomial fit of the average of largest tensor elements (green dashed line).
\label{fig:gseXXZDelta0}}
\end{figure*}
From the ground state energy the nearest neighbor spin-spin correlator $\langle S_z^i  S_z^{i+1} \rangle$ (Fig.~\ref{fig:gseXXZ}(b)) can be evaluated  as a function
of $\Delta$. The numerical result reproduces the jump at the critical point $\Delta=-1$ and the rather smooth $\Delta$ dependence at
$\Delta=1$. Above $\Delta=1$, in the AFM phase, a small `bump' in the magnetization can be identified, which our numerical analysis is able to reproduce precisely.

In Fig.~\ref{fig:gseXXZ}(c) we plot the low energy spectrum as a function of $\Delta$ with respect to the ground state energy as determined from the fixed point tensors: for all $\Delta$ one finds a degenerate ground state,
which is a consequence of the spin reflection symmetry of the XXZ Hamiltonian for $h=0$. Numerically this degeneracy is not easy to obtain close to $\Delta=1$ in the AFM phase,
and one needs large tensor sizes here.  Of course, degeneracies are never exact due to remaining finite size effects, however, those are smaller than the resolution of the plot.
Our numerical procedure assigns spin quantum numbers $S_z$ to the ground state: $S_z=\pm N/2$ in the ferromagnetic regime, several different quantum numbers in the XY phase, and $S_z=0$ in the
AFM phase as expected from analytical considerations.

The lowest gap in the ferromagnetic regime is given by $\delta=-\Delta-1$ with quantum number $S_z=\pm (N/2-1)$.
This result may be obtained by spin wave theory~\cite{MikKol} or Bethe ansatz~\cite{Albertini_1995}. Numerically we reproduce this finding quantitatively and find this state to be highly degenerate. Moreover we clearly identify a second gap with quantum number $S_z=\pm (N/2-2)$.
Both gaps close at the critical point $\Delta=-1$ and remain closed in the whole XY phase.  The gap opens again at $\Delta=1$, however, we see a completely
different $\Delta$ dependence here as compared to the ferromagnetic phase due to a Kosterlitz-Thouless phase transition into the AFM phase. In fact, close to the critical point
the gap shows a nonanalytic dependence on $\Delta$,
\begin{equation}
\delta=4\pi\exp{-\frac{\pi^2}{2\sqrt{2}} \frac{1}{\sqrt{\Delta-1}}},
\end{equation}
as obtained from Eq.~(\ref{hc})~\cite{PhysRev.151.258}.

Numerically one finds this gap to be highly degenerate with a triplet of quantum numbers $S_z=-1,0,1$  for all $\Delta>1$. This may be qualitatively understood in the limit $\Delta\rightarrow \infty$ as excitations from the Neel state~\cite{MikKol}. Moreover, one finds a second highly degenerate state with quantum numbers $S_z=\pm2$.
All these states merge at $\Delta=1$ into the critical disordered ground state of the XY phase. This analysis illustrates that the rather intricate low-lying spectrum of the XXZ model can be obtained rather precisely  from the  fixed point tensors.

It was proposed by Gu and Wen~\cite{GU2009} that the structure of fixed point tensor as a function of the control parameter may directly reveal the phase structure
of a given Hamiltonian. Here, we follow up on this proposal by plotting the largest elements of the fixed point tensors as shown in Fig.~\ref{fig:gseXXZ}(d), where
the largest tensor element is always normalized to 1.
One finds that the fixed-point tensors $T$ always have at least two non-zero elements which are numerically close or equal to 1 as seen in Fig.~\ref{fig:gseXXZ}(d).
This corresponds to the fact that the ground state is always at least doubly degenerate in the thermodynamic limit.

Since our results are calculated using a U(1) symmetric tensor representation, each element of the fixed point tensor
can be labeled by four U(1) quantum numbers.
The largest elements are found with completely different quantum numbers in different quantum phases. In fact, in the FM phase one finds just two nonzero elements with quantum numbers $(\pm N/2,0,\pm N/2,0)$. All other tensor entries are small. In the XY phase the tensor consists of many  elements close to 1 with different quantum numbers. This reflects the fact that the phase is critical. The three largest elements have quantum numbers $(0,0,0,0)$ and $(\pm 1,0,\pm 1,0)$, respectively.
Finally, in the AFM phase the two largest elements have quantum numbers $(0,0,0,0)$ at sufficiently large system sizes corresponding to two states with $S_z=0$. However, in the AFM one finds  tensor elements smaller than one but essentially non-zero with quantum numbers $(0,0,0,0)$ and $(\pm 1,0, \pm 1, 0)$.
This illustrates that the tensor structure is fundamentally different in the different phases, and
we conclude from the results shown in Fig.~\ref{fig:gseXXZ}(d) that the phase structure can indeed be immediately read off from the tensor structure. The critical points can be identified
precisely.

We now start the analysis of results for nonzero magnetic field $h$ at a few fixed values of $\Delta$. Due to the symmetry in parameter space $E(-h)=E(h)$ and $S_z(-h)=-S_z(h)$ we
only need to consider $h>0$.
At $\Delta=0$ we obtain the results shown in Fig.~\ref{fig:gseXXZDelta0}.
For small magnetic fields the system is in the XY phase, and it shows an Ising-like phase transition to the FM phase at $h_s=1$.
This Ising-like phase transition is easily recognized from the magnetization plotted in Fig.~\ref{fig:gseXXZDelta0}(b). It has a cusp at $h=1$ where the ground state becomes fully polarized with $S_z=N/2$. The numerical results agree precisely with Bethe Ansatz predictions.
The low lying spectrum is plotted in Fig.~\ref{fig:gseXXZDelta0}(c). Obviously the system is gapless in the XY phase, and the gap opens at the critical point $h_s=1$. In the ferromagnetic phase the lowest gap is given by $\delta=h-1$ with quantum number $S_z=N/2-1$.
Our calculation also clearly identifies the second gap with quantum number $S_z=N/2-2$.

The phase structure may again be determined just by plotting the largest tensor elements as done in Fig.~\ref{fig:gseXXZDelta0}(d). The largest tensor elements fluctuate as a function of the magnetic field, but still the critical point at $h_s=1$ can be extracted as follows. The average of second and third largest element is calculated and fitted as a polynomial function of up to $h^4$. The inflection point of this function coincides with phase boundary $h_s$ with absolute error 0.1. While in the XY phase one finds several tensor elements with absolute value close to one, in the FM phase there is only one such element. Furthermore, the quantum numbers of the leading elements differ significantly between the two phases ($S_z=N/2$ in the FM phase and in the range from 0 to $N/2$ in the XY phase).

\begin{figure*}
\unitlength1cm
\begin{picture}(18,9)(0,0)
 \put(1,4.5)  {\includegraphics[width=6.5cm]{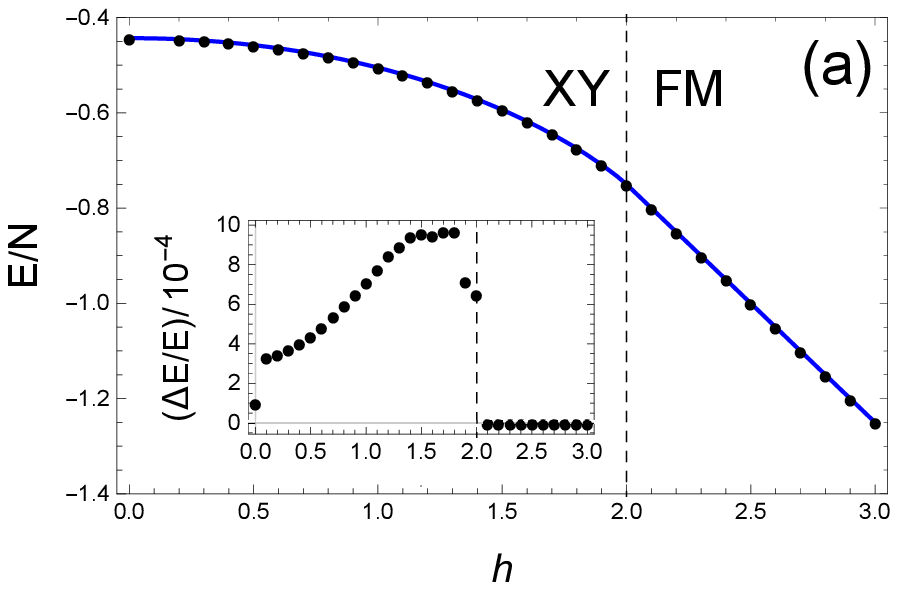}}
 \put(1,0)    {\includegraphics[width=6.5cm]{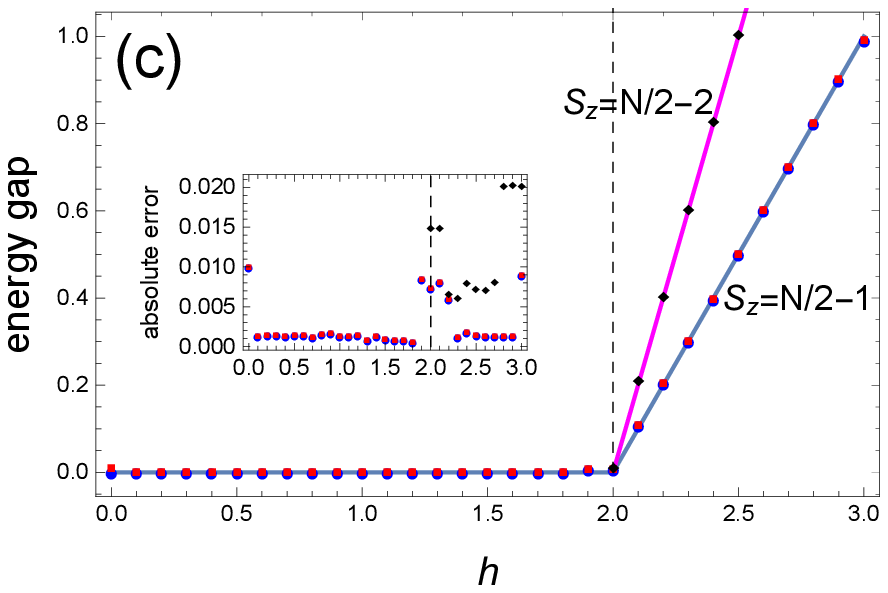}}
 \put(9,4.5)  {\includegraphics[width=6.5cm]{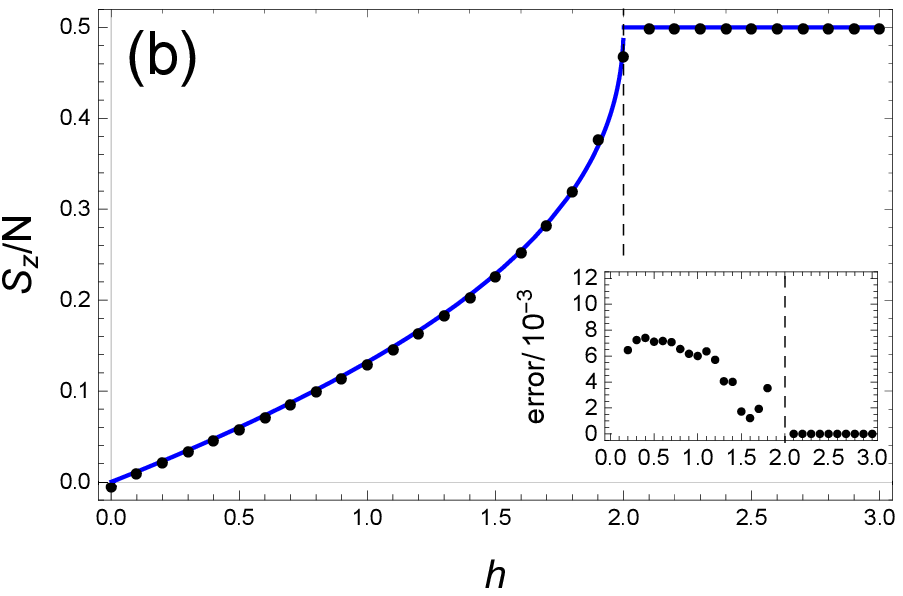}}
 \put(9,0)    {\includegraphics[width=6.5cm]{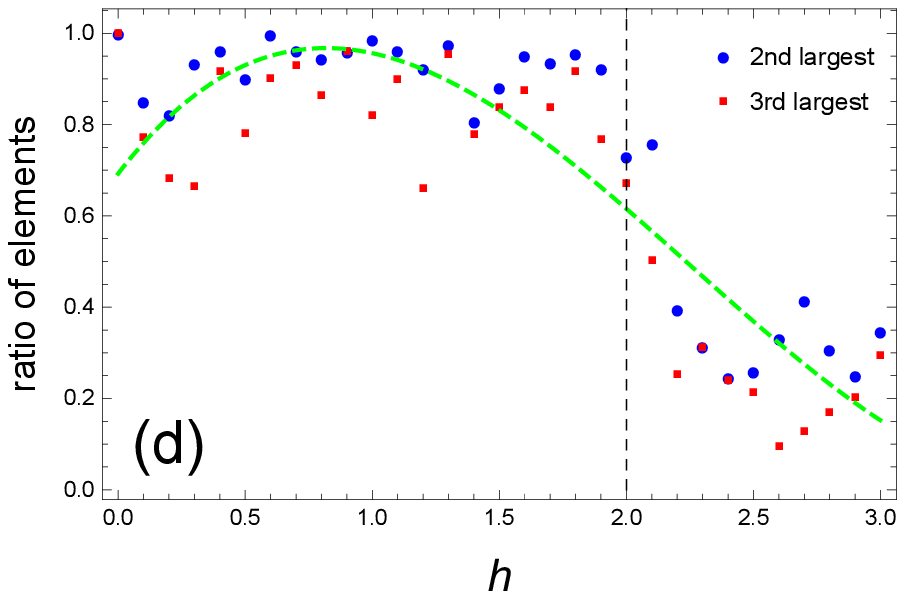}}
\end{picture}
\caption{\footnotesize
   Same Fig.~\ref{fig:gseXXZDelta0} but for $\Delta=1$.  One clearly identifies two phases (XY and FM) with a second order phase
transition at  $h_{s}=2$.
\label{fig:gseXXZDelta1}}
\end{figure*}

Let us continue our analysis at $\Delta=1$ with results shown in Fig.~\ref{fig:gseXXZDelta1}.  At this $\Delta$ the Ising-like phase transition to the FM phase occurs at $h_s=2$, but in general
the structure seen is very similar to the results obtained at $\Delta=0$. We just note that the numerical procedure is capable to capture this structure surprisingly well.

\begin{figure*}
\unitlength1cm
\begin{picture}(18,9)(0,0)
 \put(1,4.5)  {\includegraphics[width=6.5cm]{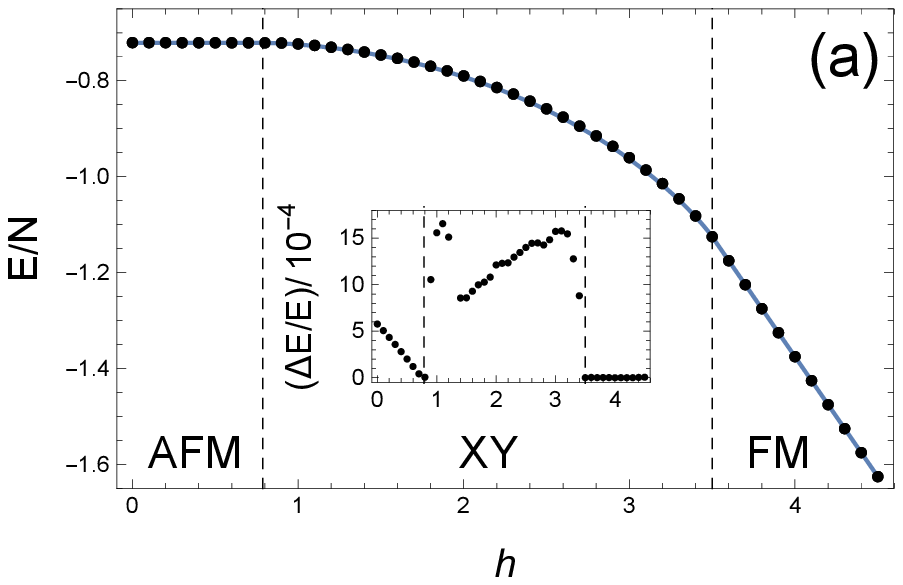}}
 \put(1,0)  {\includegraphics[width=6.5cm]{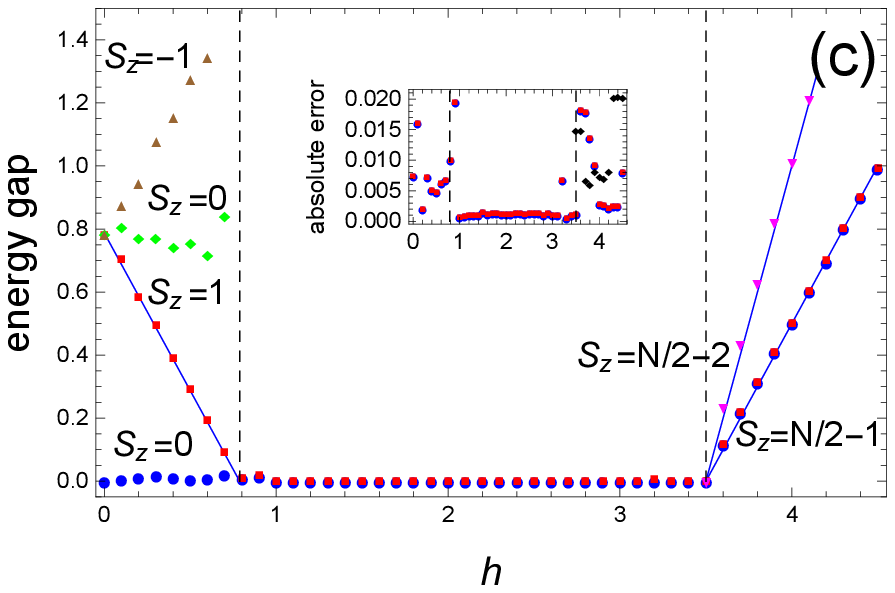}}
 \put(9,4.5)  {\includegraphics[width=6.5cm]{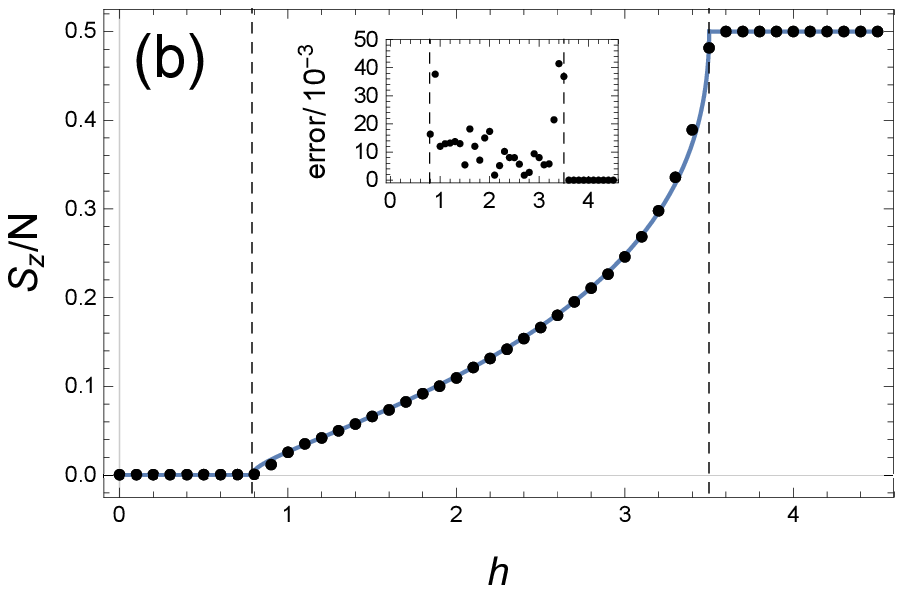}}
 \put(9,0)  {\includegraphics[width=6.5cm]{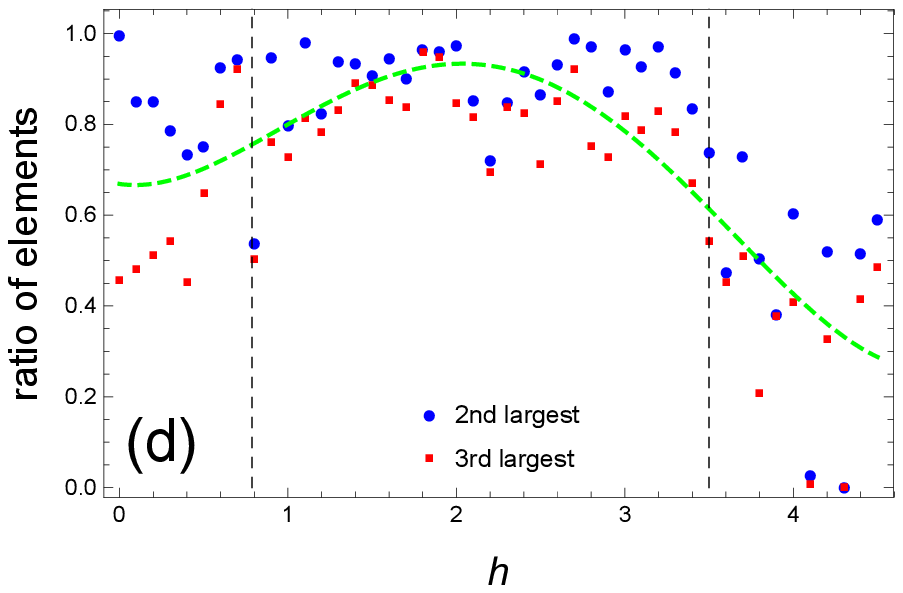}}
\end{picture}
\caption{\footnotesize Same as Fig.~\ref{fig:gseXXZDelta0} but for $\Delta=2.5$. One clearly identifies three phases (AFM, XY, FM) with second order phase
transitions at $h_{c}\approx 0.787$ and $h_{s}=3.5$. In the AFM phase we indicate the splitting of the `triplet state' by the magnetic field. Above the $S_z=1$ but below the $S_z=0$ and $S_z=-1$ triplet state there are states with larger $S_z$, which are not shown. Both phase boundaries can be obtained as inflection points of a polynomial fit of the average of largest tensor elements (green dashed line).
\label{fig:gseXXZDelta25}}
\end{figure*}

A significantly different structure for the results is observed at $\Delta=2.5$:
All three phases of the XXZ model are clearly identified by the tensor method.
This is easily seen
from the magnetization shown in Fig.~\ref{fig:gseXXZDelta25}(b) as well as the lowest lying spectrum (Fig.~\ref{fig:gseXXZDelta25}(c)).
Ground state energy and magnetization agree quantitatively with Bethe ansatz results numerically obtained from the integral equation
given Ref.~\cite{PhysRev.150.327}: The magnetization is zero to a high precision in the AFM phase, however the
vertical slope at the critical point $h_c\simeq 0.787$ is hard to obtain numerically. The magnetization increases  until the system is fully magnetized in the FM phase and again shows
a vertical slope at the critical point.

In ~\ref{fig:gseXXZDelta25}(c) we also indicate the `triplet' excited state at $h=0$, which was shown already in Fig.~\ref{fig:gseXXZ}. This state is split by the magnetic field as depicted in the figure. The $S_z=1$ state is the lowest throughout the AFM phase and joins the lowest state in the XY phase at the critical point $h_c\approx 0.787$. The $S_z=0$ and $S_z=-1$ components of the triplet state
can be easily tracked within the AFM phase, however it must be noted that there are states with larger $S_z$ below these states which are not shown. Moreover, as these states are lying high
up in the spectrum they cannot be tracked through the XY phase in our calculation.

The spectrum remains degenerate as a function of the magnetic field in the AFM and XY phase. In the ferromagnetic phase we obtain the lowest gap  $\delta=h-h_s$, where $h_s$ is given by Eq.~(\ref{hs}), while in the antiferromagnetic phase one finds a doubly degenerate ground state and a lowest gap $\delta=h_c-h$, where $h_c$ is given by Eq.~(\ref{hc}).

The behavior of largest tensor elements is shown in Fig.~\ref{fig:gseXXZDelta25}(d). The figure indicates two leading elements in the AFM phase, one leading element in the FM phase and several in the XY phase. The quantum numbers assigned to the leading elements in the XY and FM phase are the same as for $\Delta=0$ and $\Delta=1$. In the AFM phase the two leading elements have quantum numbers $(0,0,0,0)$ while the next two have quantum numbers $(\pm 1,0,\pm 1,0)$. From these observations the phase boundaries are easily determined. The average of second and third largest element is calculated and fitted again. It has two inflection points now, which differ from analytic values of $h_c$ and $h_s$ by at most 0.1.

\section{Conclusions}

The purpose of this paper is to provide a stringent quantitative test for the HOTRG tensor renormalization method
on the basis of the spin-1/2 Heisenberg XXZ model. To this end we implement this method for U(1) symmetric tensors on the basis
of a new U(1) tensor representation and its algebra. A comparison of the numerical resources required using the U(1) symmetric implementation
with a non-symmetric implementation is shown in Table~\ref{resources}.

Several previous papers concentrated on the Ising model~\cite{XIE2012,PhysRevB.89.075116} or phase boundaries of spin 1 models~\cite{GU2009}. The XXZ model has a rich phase structure including a
extended critical phase. We find that the numerics quantitatively reproduces Bethe ansatz results for ground state properties as well as the low lying spectrum
in a large parameter space. This encourages applications of the HOTRG method to systems with spins larger than 1/2 where no analytical results are available.
Such studies have been done previously by exact diagonalization with finite size extrapolations~\cite{PhysRevB.67.104401}, and we will present results obtained
using the methods presented here in a forthcoming publication.
\begin{table}\label{resources}
\begin{tabular}{|c|c|c|c|}
\hline
$m$ & time ratio & RAM, n.s. & RAM, U(1) \\
\hline
50 & 1.75 & 9.64 & 1.60 \\
\hline
60 & 1.92 & 23.7 & 4.44 \\
\hline
70 & 3.34 & 51.0 & 9.29 \\
\hline
90 & 8.42 & 178 & 25.3 \\
\hline
120 & n.a. & n.a. & 100 \\
\hline
\end{tabular}
\caption{\footnotesize CPU time ratio of non-symmetric and U(1) HOTRG, and RAM requirements (in gigabytes) for both methods, for calculations on spin-1/2 XXZ model at $\Delta=h=0$ with various tensor sizes $m$. ``n.s.'' means ``no symmetry''. ``n.a'' means that calculation was not possible on available computers.\label{resources}}
\end{table}

\acknowledgments

M.V.R. acknowledges funding by the Deutsche Forschungsgemeinschaft (DFG, German Research Foundation) under Germany's Excellence Strategy – EXC-2123/1.

\appendix

\section{Bethe Ansatz results\label{appendix-Bethe}}

The critical lines $h(\Delta)$ which separate the phases shown in Fig.~\ref{fig:phaseDiagram} are given by~\cite{PhysRev.151.258}
\begin{eqnarray}
&&h_s=1+\Delta,\label{hs}\\
&&h_c=\frac{\pi\sinh\lambda}{\lambda} \, \sum_{n=-\infty}^{\infty} \, {\rm sech} \, \frac{\pi^2}{2\lambda} (2n+1)\label{hc}
\end{eqnarray}
with $\lambda={\rm arccosh} \, \Delta$.

For $h=0$ the ground state energy per site as a function of $\Delta$  is obtained as~\cite{PhysRev.147.303, PhysRev.150.327}
\begin{eqnarray}
E/N&=&\frac{\Delta}{4} ~~~~~~~~~~~~~~~~~~~~~~~~~~~~ {\rm for }~~  \Delta\leq-1, \\
E/N&=&\frac{\Delta}{4}-\frac{1}{2}(1-\Delta^2)\times ~~~~~~~{\rm for }~~\Delta > -1 \label{e0Bethe}\\
   & & ~~\int\limits_{-\infty}^{\infty}\frac{dx}{\cosh \pi x (\cosh (2x\arccos\Delta)-\Delta)}. \nonumber
 \end{eqnarray}
At $\Delta=1$ one obtains $E/N=1/4-\log 2$.

For $\Delta=0$ one finds as a function of the magnetic field~\cite{PhysRev.150.321}
\begin{eqnarray}
E/N&=&-\frac{1}{\pi}(\sqrt{1-h^2}+h\arcsin h), ~ h \leq 1, \\
E/N&=&-\frac{h}{2}, ~~~~~~~~~~~ h > 1.
\end{eqnarray}

\section{U1Tensors}\label{U1Tensors}

A symmetric tensor network must be represented as a directed graph~\cite{PhysRevA.82.050301}, and consequently
one distinguishes incoming and outgoing indices for each tensor (similarly as one distinguishes covariant and contravariant
indices in relativity theory). In Fig.~\ref{fig:HOTRG} the direction of each edge is indicated by an arrow.
In a symmetric tensor network the contractions are always over one outgoing and one incoming index.
For mathematical details we refer to Refs.~\cite{PhysRevA.82.050301,SIN11} and references therein.

Generally
it holds that a symmetric tensor decomposes into a structural part and a degeneracy part, where the structural
part is determined by the symmetry. For U(1) symmetry this structure is particularly simple~\cite{SIN11}.
Consequently, each index $i_j$ of a U(1) symmetric tensor $T$ decomposes into a U(1) spin index $s_j$ and a degeneracy index $t_j$,
$i_j=(s_j,t_j)$, where each index belongs either to the set of incoming indices $I$ or to the set of outgoing indices $O$. Then
it holds that
\begin{equation}\label{symTenU1}
(T)_{i_1 i_2 \ldots i_t}= \delta^{s_1\ldots,s_t}_{S_{\rm in}, S_{\rm out}} P_{s_1\ldots,s_t}^{t_1\ldots t_t}
\end{equation}
with $S_{\rm in}=\sum_{I} s_j$ and $S_{\rm out}=\sum_{O} s_j$.
The  $\delta$ tensor defines the structural tensor and implements spin conservation, i.e., elements which satisfy ${S_{\rm in}= S_{\rm out}}$
are 1 and all the others are 0. The two tensors $\delta$ and $P$ are multiplied element-wise to obtain the tensor $T$.

For example, a U(1) tensor with one incoming index $i_1$ and one outgoing index $i_2$ which both run over the spin quantum numbers {-1,0,1}
must be a $3\times3$ matrix with degeneracy matrices $P^{t_1,t_2}_{-1,-1}, P^{t_1,t_2}_{0,0}, P^{t_1,t_2}_{1,1}$ as diagonal elements and 0 as off diagonal elements. The dimensions of the degeneracy matrices $t_1$ and $t_2$ are not determined by the symmetry.
Analogously, a U(1) tensor with two incoming indices would have degeneracy tensors in the anti-diagonal and 0 otherwise.

Obviously, U(1) tensors are not arrays which are typically implemented by standard computer languages.
There are various ways to deal with this problem: One could replace each 0 element by a corresponding 0-tensor and
than use standard sparse array techniques which are provided by many computer languages. However, such generic sparse
array representations are not specific for U1 tensors and therefore may lead to large computational overhead.
Here we chose to implement a specific data structure consistent with Eq.~(\ref{symTenU1}) and  implemented
the corresponding algebra in order to contract, fuse, and rearrange  such tensors (e.g. index splitting). This reduces
significantly the required computational resources and enables to handle rather large tensors, since only
the degeneracy tensors are actually stored in memory. Moreover, contractions only involve degeneracy tensors.
A similar program for SU(2) was described in Ref.~\cite{Schmoll_2018}.

Technical details of the use of such tensors in an implementation of the coarse-graining procedure described in this paper
are given in the following Appendix~\ref{U1details}.
Here, we just mention that we start
the calculation by representing the exponential of the Hamiltonian as a U1 tensor as described above, and
then use the developed tensor algebra to implement the coarse graining procedure as outlined in Fig.~\ref{fig:HOTRG}.

\section{Implementation details}\label{U1details}

The first step in order to implement a  realization of the coarse-graining procedure for the calculation of the partition sum~(\ref{part func})
as described in section~\ref{sec-ten} of this paper is the calculation of
a suitable initial 4-index tensor corresponding to the black boxes in Fig.~\ref{fig:HOTRG}(a). In order to do this one starts from the
matrix
\begin{equation}\label{t-tensor}
E_{(m_1 m_2),(m_1^\prime m_2^\prime)}=\langle m_1, m_2 | \exp( -\beta H_{12}) | m_1^\prime, m_2^\prime \rangle
\end{equation}
with $m_1, m_2, m_1^\prime, m_2^\prime$ taking the values $-s, \ldots, +s$. $H_{12}$ is the interaction Hamiltonian
between two spins, e.g. the XXZ Hamiltonian given in Eq.~(\ref{XXZham}). We calculate the matrix exponential numerically
using a series expansion and transform this matrix into a U1Tensor as defined in Eq.~(\ref{symTenU1}).
We may choose the indices $m_1$ and $m_2$ as incoming and  $m_1^\prime$ and $m_2^\prime$ as outgoing. For $s=1/2$ the structural tensor
then has the form
\begin{equation}
\delta_{S_{\rm in}, S_{\rm out}}=
\left(
\begin{array}{cc}
 \left(
\begin{array}{cc}
 1 & 0 \\
 0 & 1 \\
\end{array}
\right) & \left(
\begin{array}{cc}
 0 & 0 \\
 1 & 0 \\
\end{array}
\right) \\
 \left(
\begin{array}{cc}
 0 & 1 \\
 0 & 0 \\
\end{array}
\right) & \left(
\begin{array}{cc}
 1 & 0 \\
 0 & 1 \\
\end{array}
\right) \\
\end{array}
\right)
\end{equation}
However, only the degeneracy tensors associated with each 1 element in the tensors must be stored and handled explicitly. For the initial
tensor, of course, the degeneracy tensors are  $1\times 1\times 1\times 1$ arrays, which are easily obtained from the expression~(\ref{t-tensor}).
\begin{figure}
\unitlength1cm
\begin{picture}(0,4)(4,0)
 \put(0,0)  {\includegraphics[width=6.5cm]{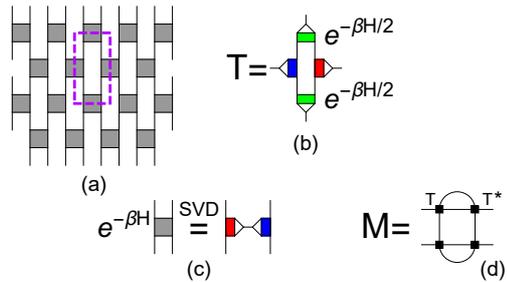}}
\end{picture}
\caption{\footnotesize
This figure is supporting the detailed discussion in the text.
(a) original tensor network, (b) initial tensor $T$, (c) left and right tensors for (b), (d) tensor $M$ for calculation of projector.
\label{fig:appendix3}}
\end{figure}

Unfortunately, the tensor corresponding to the matrix $E$ is not suitable as initial tensor $T$ for our purposes. Using this tensor as initial tensor would
yield the tensor network shown in Fig.~\ref{fig:appendix3}(a), which is obviously not compatible with the structure shown in Fig.~\ref{fig:HOTRG}(a). However, it is easily possible
to transform the tensor network shown in Fig.~\ref{fig:appendix3}(a) into the required form. Such a transformation is described e.g. in Ref.~\cite{PhysRevB.95.045117}. Our variant is shown in Fig.~\ref{fig:appendix3}(b). The top and bottom tensors are obtained from a U1Tensor representation of the matrix $\exp(-\beta H/2)$ while
the left and right tensors are obtained (Fig.~\ref{fig:appendix3}(c)) from a singular value decomposition of the matrix~(\ref{t-tensor}), both transformed into U1Tensors as described above.
Then, in order to calculate the tensor shown in Fig.~\ref{fig:appendix3}(b), one needs to contract and fuse four U1Tensors, such that the  U1 structure as defined in Eq.~(\ref{symTenU1}) is preserved.

Obviously, fusing two U1 spin-1/2 indices $s_1$ and $s_2$, which take values $-1/2, 1/2$, yields a U1 spin-1 index taking values $-1, 0, 1$,
with the 0 index  two-fold degenerate in line with the fact that we obtain spin 0 in two different ways. This basic fact must be implemented in a general
way for U1 indices such that the U1 structure of a tensor and its index labels is correctly maintained during the calculation. Specifically, the tensor shown in Fig.~\ref{fig:appendix3}(b) is labeled in all
four dimensions by spin-1 indices $-1,0,1$ with a twofold degeneracy for each 0 index. It is exactly this tensor which serves a suitable initial tensor $T$
for our implementation as indicated by the black box in Fig.~\ref{fig:HOTRG}(a). We emphasize that all calculations in order to obtain this tensor are essentially exact
(up to numerical rounding of machine precision numbers). Degeneracy tensors are not truncated at this stage.

The next step is to contract two of such basic tensors $T$ as indicated in~Fig.~\ref{fig:HOTRG}(b). This is done in such a way that only degeneracy tensors must be contracted
and stored. Initially the two left and right indices of this tensor are just fused according to the procedure described above, and we obtain spin-2 indices taking the values  $-2,-1,0,1,2$
with degeneracies $1,4,6,4,1$, respectively. These degeneracy indices are not truncated, and the resulting tensor network is still representing exactly the partition sum to be calculated.
We continue such contractions and fusions until the total number of degeneracy indices exceeds a predetermined number $m$. When that happens we start to truncate the resulting tensor
in the manner indicated in Fig.~\ref{fig:HOTRG}(c).

To this end we determine a projector $P$ which is symbolized by the triangle in  Fig.~\ref{fig:HOTRG}(c). This projector is determined using
the higher order singular value decomposition (HOSVD). The key principles of HOSVD are described in Ref.~\cite{Lat_2000} and are reviewed in Ref.~\cite{XIE2012}. Here we do not
review mathematical details of this method  but just state how the calculation of the projector is implemented following the graphical representation in Fig.\ref{fig:appendix3}(d). The special requirement we have here is that the U1 structure is maintained.
As indicated in Fig.\ref{fig:appendix3}(d), we determine a tensor $M$ which we fuse into a block diagonal matrix.  For each block we determine eigenvectors and eigenvalues and select from each block those
eigenvectors corresponding to the largest of {\it all}  eigenvalues until just $m$ eigenvectors remain. In this way the size of {\it one} index of several eigenvector blocks reduces (some blocks may disappear entirely, a few may stay unaffected by this selection).
After the selection we reconstruct a three-index U1Tensor $P$ by splitting the unreduced index of each eigenvector block into the original U1 index structure. In this way we obtain
a three-index U1Tensor which serves as a suitable projector according to HOSVD theory. The projected U1tensor which is depicted as an open box in Fig.~\ref{fig:HOTRG}(d) serves as the building block
of the coarse-grained tensor network as shown in Fig.~\ref{fig:HOTRG}(e).

The course-graining procedure is then repeated both in imaginary time direction and in space direction until we reach a fixed point tensor.
From this fixed point tensor all observables are calculated, in particular the low lying spectrum.


\begin{thebibliography}{10}

\bibitem{MikKol}
H.-J. Mikeska and A.K. Kolezhuk.
\newblock One-dimensional magnetism.
\newblock {\em Lect. Notes Phys.}, 645:1--83, 2004.

\bibitem{WHI93a}
Steven~R. White and David~A. Huse.
\newblock Numerical renormalization-group study of low-lying eigenstates of the
  antiferromagnetic {S=1 Heisenberg} chain.
\newblock {\em Phys. Rev. B}, 48(6):3844--3852, Aug 1993.

\bibitem{WHI93b}
Steven~R. White.
\newblock Density-matrix algorithms for quantum renormalization groups.
\newblock {\em Phys. Rev. B}, 48(14):10345--10356, Oct 1993.

\bibitem{LEV2007}
Michael Levin and Cody~P. Nave.
\newblock Tensor renormalization group approach to two-dimensional classical
  lattice models.
\newblock {\em Phys. Rev. Lett.}, 99:120601, Sep 2007.

\bibitem{GU2009}
Zheng-Cheng Gu and Xiao-Gang Wen.
\newblock Tensor-entanglement-filtering renormalization approach and
  symmetry-protected topological order.
\newblock {\em Phys. Rev. B}, 80:155131, Oct 2009.

\bibitem{Kadanoff_1966}
L.~P. Kadanoff.
\newblock Scaling laws for ising models near {${T}_{c}$}.
\newblock {\em Physics}, 2:263--272, Jun 1966.

\bibitem{XIE2012}
Z.~Y. Xie, J.~Chen, M.~P. Qin, J.~W. Zhu, L.~P. Yang, and T.~Xiang.
\newblock Coarse-graining renormalization by higher-order singular value
  decomposition.
\newblock {\em Phys. Rev. B}, 86:045139, Jul 2012.

\bibitem{PhysRevLett.115.180405}
G.~Evenbly and G.~Vidal.
\newblock Tensor network renormalization.
\newblock {\em Phys. Rev. Lett.}, 115:180405, Oct 2015.

\bibitem{PhysRevLett.115.200401}
G.~Evenbly and G.~Vidal.
\newblock Tensor network renormalization yields the multiscale entanglement
  renormalization ansatz.
\newblock {\em Phys. Rev. Lett.}, 115:200401, Nov 2015.

\bibitem{PhysRevLett.118.110504}
Shuo Yang, Zheng-Cheng Gu, and Xiao-Gang Wen.
\newblock Loop optimization for tensor network renormalization.
\newblock {\em Phys. Rev. Lett.}, 118:110504, Mar 2017.

\bibitem{PhysRevB.89.075116}
Hiroshi Ueda, Kouichi Okunishi, and Tomotoshi Nishino.
\newblock Doubling of entanglement spectrum in tensor renormalization group.
\newblock {\em Phys. Rev. B}, 89:075116, Feb 2014.

\bibitem{SIN11}
Sukhwinder Singh, Robert N.~C. Pfeifer, and Guifre Vidal.
\newblock Tensor network states and algorithms in the presence of a global
  {U(1)} symmetry.
\newblock {\em Phys. Rev. B}, 83:115125, Mar 2011.

\bibitem{PhysRevB.93.054417}
Mykhailo~V. Rakov, Michael Weyrauch, and Briiissuurs Braiorr-Orrs.
\newblock Symmetries and entanglement in the one-dimensional spin-$\frac{1}{2}$
  xxz model.
\newblock {\em Phys. Rev. B}, 93:054417, Feb 2016.

\bibitem{PhysRevB.95.045117}
G.~Evenbly.
\newblock Algorithms for tensor network renormalization.
\newblock {\em Phys. Rev. B}, 95:045117, Jan 2017.

\bibitem{Lat_2000}
L.~De Lathauer, B.~De. Moor, and J.~Vandewalle.
\newblock {\em SIAM J. Matrix Anal. Appl.}, 21:1253, May 2000.

\bibitem{PhysRevA.82.050301}
Sukhwinder Singh, Robert N.~C. Pfeifer, and Guifr\'e Vidal.
\newblock Tensor network decompositions in the presence of a global symmetry.
\newblock {\em Phys. Rev. A}, 82:050301, Nov 2010.

\bibitem{PhysRev.147.303}
C.~N. Yang and C.~P. Yang.
\newblock Ground state energy of {Heisenberg-Ising} lattice.
\newblock {\em Phys. Rev.}, 147:303--306, Dec 1965.

\bibitem{PhysRev.150.327}
C.~N. Yang and C.~P. Yang.
\newblock One-dimensional chain of anisotropic spin-spin interactions. {II.
  Properties} of the ground-state energy per lattice site for an infinite
  system.
\newblock {\em Phys. Rev.}, 150:327--339, Oct 1966.

\bibitem{PhysRev.151.258}
C.~N. Yang and C.~P. Yang.
\newblock One-dimensional chain of anisotropic spin-spin interactions. {III.
  Applications}.
\newblock {\em Phys. Rev.}, 151:258--264, Nov 1966.

\bibitem{Albertini_1995}
G.~Albertini, V.~E. Korepin, and A.~Schadschneider.
\newblock {XXZ} model as an effective hamiltonian for generalized {Hubbard}
  models with broken {$\eta$}-symmetry.
\newblock {\em J. Phys. A: Math. Gen.}, 28(10):L303--L309, May 1995.

\bibitem{PhysRevB.67.104401}
Wei Chen, Kazuo Hida, and B.~C. Sanctuary.
\newblock Ground-state phase diagram of {$S=1$ $\mathrm{XXZ}$} chains with
  uniaxial single-ion-type anisotropy.
\newblock {\em Phys. Rev. B}, 67:104401, Mar 2003.

\bibitem{PhysRev.150.321}
C.~N. Yang and C.~P. Yang.
\newblock One-dimensional chain of anisotropic spin-spin interactions. {I.
  Proof of Bethe's} hypothesis for ground state in a finite system.
\newblock {\em Phys. Rev.}, 150:321--327, Oct 1966.

\bibitem{Schmoll_2018}
P.~Schmoll, S.~Singh, M.~Rizzi, and R.~Orus.
\newblock A programming guide for tensor networks with {SU(2)} symmetry.
\newblock {\em arXiv:1809.08180v1}.

\end{thebibliography}
\end{document}